\documentstyle[aps,epsfig]{revtex}

\begin{document}
\title{Phases of Neon, Xenon, and Methane Adsorbed on Nanotube Bundles}
\author{M. Mercedes Calbi$^{a}$, Silvina M. Gatica$^{b}$, Mary J. 
Bojan$^{c}$ and Milton W. Cole$^{a}$}.
\address{$^{a}$Department of Physics, Pennsylvania State University, 
University Park, PA 16802 USA}
\address{$^{b}$Departamento de F\'{\i}sica, Facultad de Ciencias Exactas y 
Naturales, Universidad de Buenos Aires, 1428 Buenos Aires, Argentina}
\address{$^{c}$Department of Chemistry, Pennsylvania State University, 
University Park, PA 16802 USA}
\date{\today}
\maketitle

\begin{abstract}

We explore the behavior of neon, xenon, and methane films adsorbed on the 
external surface of a bundle of carbon nanotubes. The methods used are 
classical: a
ground state calculation, by grand potential energy minimization, and the
grand canonical Monte Carlo (GCMC) method of simulation. Our results are
similar to those found recently in a GCMC study of Ar and Kr. At low chemical 
potential
(pressure) the particles form a quasi-one dimensional phase within
the groove formed by two contiguous tubes. At higher chemical potential,
there occurs a "three-stripe" phase aligned parallel to the groove (except 
for xenon). This is
followed by monolayer and bilayer phases. The low temperature monolayer phase is striped; the number of stripes per nanotube is a quantized function of the adatom size. In the neon case, the bilayer 
regime also includes a second layer groove phase. Our results are compared 
with recent thermal and diffraction experiments. We find no evidence of a 
zig-zag phase reported recently.

\end{abstract}

\section{Introduction}

\vspace{.5cm}

Attention has been drawn in recent years to the problem of adsorption within 
and outside of carbon nanotubes and bundles of carbon nanotubes (ropes). This 
interest is directed towards both the  possibility of novel one-dimensional 
(1D) and quasi-1D phases \cite{rmp}, of importance to fundamental research, and to 
possible applications, such as gas storage, chemical sensing, gas 
purification and isotope separation \cite{peter,karl}. This paper is concerned with the phase 
transition behavior of gases adsorbed on the outside of a bundle of nanotubes \cite{bundle}. 
While one might think of this region as a site for 2D matter (as opposed to 
1D), the actual behavior is more varied and interesting. As discussed in a 
recent study  of this problem \cite{gat} (henceforth called I), the 
predicted phases of the adsorbate at low T are a sensitive function of 
pressure (P) or the related thermodynamic variable, the chemical potential $\mu$. 
They range from quasi-1D to 2D to 3D phases as $\mu$ is increased. Recent experiments by Talapatra and Migone \cite{mig}, Muris et al. \cite{where} have provided some quantitative support for the results of I in the cases of Ar, Xe and CH$_4$.

In I, we studied the adsorption of Ar and Kr atoms within a simplified model, 
consisting of a planar array of parallel nanotubes, like strands of pasta 
laid next to one another on a table. That model was intended to represent the 
outside surface of a perfectly ordered bundle of very large radius (much 
larger than the radius of the individual tubes). The assumption of a periodic 
geometry helps to motivate a further approximation, the use of periodic 
boundary conditions in the grand canonical Monte Carlo method, employed both 
here and in I. Evidently, these assumptions sacrifice some degree of reality; 
more accurate predictions would require a specification of the detailed 
geometry at the surface. Such complete information does not exist, to the 
best of our knowledge, about any specific bundle of tubes.

This paper extends I by presenting two new calculations. The first is 
a determination of the equation of state of adsorbed CH$_4$ and Ne at T=0. The 
second is a set of simulations of Ne and Xe at various temperatures. The resulting phase 
behavior does not differ qualitatively from that reported in I, except in some respects associated with the varying diameter of the adsorbate.

Figure 1 depicts the potential energy V(r) experienced by a CH$_4$ molecule 
in the 
vicinity of its equilibrium position above the surface. The x-y plane (shown) 
is perpendicular to the tubes and the origin is taken to be the position of 
the potential energy minimum. As in most of our previous studies, including 
I, V has been computed by summing semiempirical Ne-C interactions. 
Furthermore, the atomic structure of the tubes is ignored.  This assumption 
eliminates the possibility of commensurate phases, arguably the most 
significant limitation of our work \cite{fn}. In the absence of knowledge about the 
chirality and relative registry of adjacent tubes, however, it seems prudent 
to employ this assumption.

The data in Fig. 1 reveal a very narrow region of strongly attractive 
potential energy. We refer to this region, oriented parallel to the z axis, 
as a ``groove''. It is not surprising that at low $\mu$ the adsorbate phase
 is a 1D fluid (or crystal at T=0), in which the molecular motion 
parallel to 
the groove is free. In this phase, the  density is concentrated very close to 
the z axis. Fig. 2 depicts two hypothetical phases which might occur at 
higher values of $\mu$, after the groove becomes saturated. We refer to these 
as 
the ``zig-zag'' and ``3-stripe'' phases, respectively. In I,
calculations for 
Ar and Kr yielded evidence of a very small range of chemical potential for 
which the 3-stripe phase occurs but found no evidence of a zig-zag phase. 
Recent neutron scattering studies of CD$_4$, however, yielded data which were 
interpreted as providing evidence of a zig-zag phase \cite{michel}. This apparently 
contradictory situation has provided one motivation for the present study. A second motivation is to extend I to 
include smaller (Ne) and larger (Xe) atoms in order to discern the effect of 
size relative to the distance between grooves, measured across the tube's surface.

In Section 2 of this paper, we describe ground state calculations of the 
phase behavior of CH$_4$ and Ne. Section 3 presents the results of simulations 
of Ne and Xe adsorption at finite temperatures. In no case is there found evidence of a stable zig-zag phase. 

\section{Ground State Calculation}

	We consider three possible submonolayer phases of 
the adsorbed CH$_4$ gas on the external surface (Fig. 2): A linear phase, a zig-zag 
phase, and a 3-stripe phase. Our goal in this section is to determine the evolution of structure as a function $\mu$ at $T=0$. To find out when these phases occur as the number of 
adsorbed atoms per unit length $\rho=N/L$ increases, we minimize the grand 
potential energy $\Omega=E-\mu N$ as a function of $\rho$. $E$ is the total energy of 
the particles 

\begin{equation}
E(x,y;\rho)= \sum_{i} V(x_i,y_i)+ \frac{1}{2} \sum_{i,j} 
U_{LJ}(r_{ij})
\end{equation}

This expression includes the potential energy $V(x,y)$ at the external surface of the bundle 
plus the mutual interaction energy. $V(x,y)$ is obtained by
summing the potential energy contributions of the adjacent nanotubes \cite{stan}, and $U_{LJ}$ is a 
Lennard-Jones pair potential representing the interaction 
between molecules. Our tubes are modeled as continuum sheets of carbon, obviating any consideration of commensurate phases. Our neglect of quantum effects is a similar simplification \cite{neon}, as is the neglect of anisotropy in the gas-carbon and intermolecular potentials \cite{aniso}.

For a given density, the equilibrium configuration is obtained by minimizing
 the total energy with respect to the position of the particles in the 
(x,y) plane. At low density, the first term in Eq.(1) is minimized if the 
molecules reside 
at the potential energy minimum position (the groove); the ground state
corresponds to a 1D spacing $\Delta z$ = 4.17 \AA\, and a CH$_4$-CH$_4$ interaction energy per particle $E_{int}/N= -167$ K. The minimum substrate potential energy for CH$_4$ (-2019 K) is approximately twice the potential energy minimum on a single graphene sheet (-1070 K), the ratio differing from two because of the tube's curvature. At higher density, a zig-zag phase becomes possible, in principle, in order to minimize the mutual repulsion. Minimizing (1) yields a zig-zag structure with particles at (x,y) values shown in figure 3. These positions lie on the curve provided in figure 4; the points correspond to minima of $V(x,y)$ with respect to $y$ for each $x$.

The resulting energies are seen in figures 5 and 6. Note that the mutual interaction in the linear phase is rapidly rising as a function of $\rho$ near $\rho=0.28$ \AA$^{-1}$. The hypothetical transition to a zig-zag phase yields a marked reduction in this energy, while $V$ increases due to particles' moving away from the most favored position (0,0). Even though the individual contributions to Eq.(1) change discontinuously at the transition, the total energy is continuous there.

Figure 6 shows the extreme sensitivity of the total energy as a function of the displacement for the zig-zag structure. Note that for $\rho < 0.285$ \AA$^{-1}$ the most favored structure occurs at $x=0$ (i.e. a linear phase). This energy increases rapidly until $\rho=0.285$ \AA$^{-1}$, at which point a transition occurs to the positions (x,y) $\approx$ (0.8 \AA, 0.8 \AA). 

We consider now the other candidate for a submonolayer phase, the 3-stripe phase depicted in figure 2. In contrast to the behavior for the zig-zag transition, the energy and positions of the 3-stripe phase are smooth functions of $\rho$, as seen in figure 7. In this case, the interaction energy decreases (from a nonzero starting 
value due to the transverse interaction between the groove molecules and the two outermost lines of particles) while the 
potential energy per particle remains almost constant; thus the particles self-organize in 
equilibrium configurations determined mainly by the mutual interaction energy.
 
In order to determine the ground state energy of the system and  
the spectrum of transitions between the phases, we minimize the 
grand potential energy $\Omega$ as a function of the density; this is 
equivalent to a Maxwell construction for $\mu(\rho)$. 
At the transition, the equilibrium conditions are
the equality of the chemical potential and the 1D pressure ($-\Omega/L$) 
of the two coexisting phases. Figure 8 presents the resulting adsorption 
isotherm (T=0). The stable low density 
phase is linear, with lattice constant between 
4.17 \AA \, and 4.00 \AA \,($\rho=0.24$ to $0.25$ \AA$^{-1}$). When this phase 
is compressed above 0.25 \AA$^{-1}$, a transition to the 3-stripe phase occurs 
(which has $\rho = 0.735$ \AA$^{-1}$). As seen in the figure, the 
previously discussed transition to the zig-zag phase does not occur because it lies at a much higher chemical potential and grand free energy than the 3-stripe phase.

Analogous calculations for Ne on the external surface of the bundle yield the same qualitative phase behavior. In Table II and III we summarize the results for CH$_4$ and Ne, indicating the interparticle distance between nearest neighbors in each phase. Note that the interparticle distances between the molecules in the unit cell are very similar for every stable configuration. The particles tend to form a close packed structure (1D, 2D and 3D for linear, zig-zag and 3-stripe phases, respectively) driven mainly by just the interparticle interaction because the potential energy per atom remains nearly constant over the density range in which each phase is stable (see Figures 5, 7 and 8).  
Our finding that the zig-zag phase is not stable appears to disagree with the interpretation of the experimental results of Ref.\cite{michel}. This disagreement is discussed in Section IV.

\vspace{1cm}

\section{Monte Carlo calculations}

The grand canonical Monte Carlo (GCMC) technique was used to simulate isotherms of Ne and Xe adsorbed on a bundle of nanotubes. As with the ground state study of Section II, a Lennard-Jones potential 
is used to
describe both the atom-atom and the atom-C interactions. A
more detailed description of the model and the simulation technique
is provided in I.  The potential parameters (well depths and diameters) are
presented in Table III.  

Figure 9 depicts the results of Ne GCMC simulations at three temperatures 
relevant to typical Ne adsorption experiments, T=12, 18 and 25 K. The left scale in this and subsequent figures is the number of atoms per groove per unit cell, which has a length 10 $\sigma_{gg}$. To convert this to a commonly reported measure of adsorption (mmoles/g) requires a specification of the number of grooves on the surface of a bundle relative to the mass of the bundle. Following I, we assume the adsorbent to be a hexagonal bundle containing 37 tubes and 18 grooves. In this case, the conversion involves multiplying the left scale by a factor of 0.10 (0.071) mmoles/g for Ne (Xe), the factor varying inversely with $\sigma_{gg}$. If the geometry were to differ from this, the right scale would require a multiplication by the ratio $(N_{groove}/18)(37/N_{tube})$. 

The behavior is 
similar in some, but not all, respects to that reported in I for gases which 
have larger diameters (Ar and Kr). At the lowest P, 
adsorption occurs primarily within the groove formed by the assumed close 
contact of the tubes. This phase's behavior is therefore well described by 
the 1D classical equation of state. After the groove region is filled, there 
occurs a jump in density by a factor 
of order three to the 3-stripe phase. Figure 10 shows a top view of the 
density projected onto the x-y plane. The GCMC data yield no evidence of 
the hypothetical zig-zag phase. This 
finding is consistent with both the ground state study of the previous 
section and the results in I for other gases. 

At somewhat higher P, there 
occurs a second jump to a monolayer film. Note the presence of hysteresis in 
Figure 9, indicative of two states of similar free energy. One might be called a 
7-stripe phase and the other an 8-stripe phase. Such hysteresis was not 
observed in I. Hysteresis of this kind (often of much larger magnitude) is 
commonly seen in experiments and simulations involving porous media, arising 
from capillary condensation. It is rarely seen in thin physisorbed film 
growth on flat surfaces, except in cases of surface heterogeneity \cite{cas}. 

Evidence of the two phases' stability (or metastability) is their nearly 
constant coverage  over an extended range of chemical potential and
 temperature, as seen in the isotherms. Figures 10 and 11 present  views 
of these phases' density configurations in the x-y and x-z planes. Note
 that the lower density, 7-stripe, phase includes some atoms which are not 
well
 defined in position at the low temperature (12 K) of the simulation; these
 sit at the highest potential energy positions, near the potential energy
 ridge, at the top of a nanotube. The 8-stripe phase, in contrast, has an
 extra line of atoms squeezed in there; these last lines to be added involve atoms which are very localized, as are those close to the groove. The 7-stripe phase, 
whose most weakly bound atoms are very delocalized,
 has a slightly lower (-659 K vs. -654 K) energy/atom than the higher density
 phase.  However, the equilibrium condition involves the minimum of the grand
free energy, which is the product of the coverage and the difference between
the Helmholtz free energy/atom F/N and the chemical potential of the
system. At very low temperature, the difference between F and E is small;
in 2D, it is proportional to $k_B T$ times the square of the ratio of the
lattice constant to the thermal phonon wavelength \cite{book}. At our lowest
temperature, T=12 K, this difference amounts to less than 10 K per atom and
is negligible compared to the magnitude of E/N, so we may neglect the
difference between F and E in this regime. At T=12 K and P = 10$^{-16}$ 
atm, for
example, the chemical potential is -530 K. Hence, the grand free energy is
 equal to the coverage times a negative number, of large magnitude ($E/N-\mu \approx -120$ K). The 14 \%
 larger coverage of the 8-stripe phase ensures that its grand free energy is
 lower than that of the (energetically favored) 7-stripe phase. This 
comparison is consistent with the commonly held belief that the desorption 
branch of a hysteretic isotherm is the more stable branch. As is commonly
 seen in simulations and experiment, hysteresis occurs when comparably
 energetic phases are separated by phase space jumps. Here, the jump involves
 both a fairly large coverage change and a spatial rearrangement associated
 with squeezing in the extra line of atoms. It is therefore not surprising 
to see 
this phenomenon at this first order transition.
 
The monolayer coverage in this Ne case (7 or 8-stripe) is 
consistent with that found in I for both Ar (6 stripe) which has $\approx$ 
25 \% higher diameter and Xe ($\approx$ 33 \% larger, 5-stripe). At even 
higher P, for Ne there occurs a $\approx$ 10 \% coverage step,  observed 
in Figure 10 to correspond to the appearance of a single line of atoms above the groove. 
This is interpreted as adsorption of a second linear phase localized within a 
Ne-coated groove. This second groove phase extends over a much smaller 
range of chemical potential than does the first groove filling because 
of the healing 
out of the deep potential  well near the groove, a consequence of the Ne 
layer's relatively weak attraction. We note that such a second groove phase 
is absent from the phase behavior of the larger diameter gases studied in I, as well as that of Xe, discussed below.

This absence is plausible because the monolayer films of such large gases are 
as thick as their diameters. Second layer adsorption then occurs in the 
presence of a relatively smooth (laterally) external potential. In the 
contrasting case of a very small atom, like Ne, the second layer experiences a 
residual effect of the substrate's groove attraction, as seen in the 
potential energy contours of Figure 1. Hence, the second layer groove phase occurs for Ne but not larger gases we have studied.

Finally, the discontinuous feature occurring at the highest P (10$^{-4}$ atm at 18 K) represents the 
completion of a bilayer film, manifested as an 80 \% jump in coverage close to 
saturated vapor pressure. Further pressure increase results in wetting film 
growth, as is expected in the case of the strongly attractive potential 
provided by the layer of nanotubes.

Figure 12 presents results for the simulation of Xe adsorption. The 3-stripe 
phase is absent, in contrast to the case of the smaller gases. This reflects 
the relatively large contribution to the energy of the Xe-Xe interaction, on the one hand, and the 
relatively small variation of the adsorption potential (once the groove is 
filled) due to the large size of the Xe atoms.

\vspace{1cm}

\section{Discussion and conclusions}

In this study we explored the sequential appearance of phases on the outer 
surface of a bundle of nanotubes. Ground state calculations for Ne and CH$_4$ 
found the 1D groove phase to be followed by the 3-stripe phase. The zig-zag 
phase is not stable relative to either of these at any 1D density. Although we did not simulate methane, we found no evidence of a zig-zag phase for Ne, Ar, Kr (which is very similar to methane) or Xe in any of our simulations. This 
general finding of no zig-zag phase is in disagreement with the conclusion of Ref.\cite{michel}, which deduced the 
presence of a zig-zag phase as part of an interpretation of neutron diffraction data of CD$_4$. The 1D 
lattice constant reported in the latter study (7.4 \AA) is quite inconsistent, 
incidentally, with that (4 \AA) of the metastable zig-zag phase found here, 
in any case. Indeed, a common feature of all of our results (Table II) is 
a nearest neighbor spacing near 4 \AA, close to that of both 2D and 3D phases 
of methane. This constancy is somewhat surprising in view of the deep and 
rapidly varying substrate potential shown in Fig. 1; the computed behavior 
reflects the stiffness of this quasi-2D lattice. It is difficult to identify the origin of the apparent discrepancy between experiment and theory. The principal sources of our uncertainty are assumptions concerning the form of the interactions and the geometry. We do not see how any plausible change in our model might result in the large change needed for the zig-zag phase to become favored relative to the 3-stripe phase. A study of the effect  of substrate disorder on the diffraction seems warranted; it remains our primary hope for a resolution of the 15 \% discrepancy between period predicted here for the 3-stripe phase 
and the period deduced from the very broad diffraction data.

Our simulation studies of Ne between 12, 18 and 25 K are quite consistent with 
the ground state study of that gas. The 3-stripe phase is present over a more 
extended range of chemical potential than was found for the larger gases, a 
consequence of the ability of the Ne to nestle into the groove region in 
order to lower its potential energy. At somewhat higher chemical potential, 
other phases appear. The monolayer phase exhibits hysteresis, apparently 
because there are two energetically similar monolayer densities. At slightly 
higher density, there appears a second groove phase, in which a line of atoms 
occupies second layer sites above the groove. This, too, was not seen in I 
and occurs because of the small size of Ne. At even higher chemical 
potential, the second layer forms and thick films evolve.

The results for Xe differ from those of the smaller diameter gases. The linear groove phase undergoes a transition to a 5-stripe monolayer phase at low T, followed by a jump to a full bilayer phase. This behavior is a logical concomitant of the larger diameter and relatively stronger adsorbate-adsorbate attraction.

The most logical experiments to propose are thermodynamic and neutron 
diffraction studies. Indeed, these are already being carried out for  methane 
and the larger gases. Neon, not yet studied experimentally, would be of particular interest 
because its size makes it a more sensitive probe of the attractive potential. 
Helium and hydrogen are yet to be studied theoretically; these systems would require quantum 
calculations, except at high temperatures \cite{neon}.

\section*{Aknowledgments}

We are grateful to Victor Bakaev, Michel Bienfait, Renee Diehl, Aldo Migone and Peter Zeppenfeld for discussion and communication of results prior to publication. This research was supported by the Army Research Office, the Petroleum Research Fundation of the American Chemical Society, Fundaci\'on Antorchas and CONICET.

\newpage

\begin{table}[tbh]
\caption{Components of nearest-neighbor separation vectors ${\bf {r}}_{ij}$ in each phase of CH$_4$. For the zig-zag phase ${\bf {r}}_{ij}$ is between stripes and for the 3-stripe phase it is between a groove atom and an atom in an adjacent stripe. We show the whole range of possible distances for the linear phase while for the zig-zag and 3-stripe phases we indicate the values corresponding to the onset ($\rho =$ 0.5 \AA$^{-1}$ and 0.735 \AA$^{-1}$, respectively). }
\begin{tabular}{|c|c|c|c|c|c|}
Phase &lattice constant (\AA)&$\Delta z$ (\AA)&$\Delta x$ (\AA)&$\Delta y$ (\AA)&$r_{ij}$ (\AA) \\ \hline
linear&(4.17-4.00)&(4.17-4.00)&0&0&(4.17-4.00) \\ \hline
zig-zag & 4.024  & 2.012 & 3.54 & 0 & 4.072  \\ \hline
3-stripe & 4.08 & 2.04 & 2.504 & 2.469 & 4.065 \\ \hline
\end{tabular}
\end{table}

\begin{table}[tbh]
\caption{Same as Table II, showing corresponding results for Ne. The onset values for the zig-zag and 3-stripe phases are $\rho=$ 0.68 \AA$^{-1}$ and 1.00 \AA$^{-1}$, respectively.}
\begin{tabular}{|c|c|c|c|c|}
& $\Delta z$ (\AA) & $\Delta x$ (\AA) & $\Delta y$ (\AA) &$r_{ij}$ (\AA) \\ \hline
linear & (3.08-2.92) & 0 & 0 & (3.08-2.92) \\ \hline
zig-zag & 1.47 & 2.58 & 0 & 2.97  \\ \hline
3-stripe & 1.5 & 1.63 & 1.99 & 2.98 \\ \hline
\end{tabular}
\end{table}

\begin{table}[tbh]
\caption{Lennard-Jones parameters of gas-gas and gas-carbon interaction.}
\begin{tabular}{|c|c|c|c|c|}
Gas & $\sigma_{gg}$ (\AA)& $\epsilon_{gg}$ (K) & $\sigma_{gc}$ (\AA)& $\epsilon_{gc}$ (K)\\ \hline
CH$_4$ & 3.72 & 161.35 & 3.56 & 67.2 \\ \hline
 Ne     & 2.75 & 35.6   & 3.075 & 31.6 \\ \hline 
 Xe     & 4.1 & 221 & 3.75 & 78.7 \\ \hline 
\end{tabular}
\end{table}

\begin{figure}
\centerline{\psfig{file=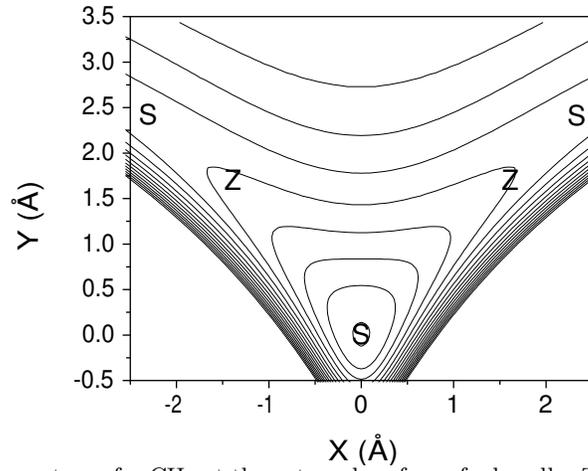, width=9cm}}
\caption{Equipotential energy contours for CH$_4$ at the external surface of a bundle. The contours correspond to energy values from -2200 K to 800 K in 200 K steps. The letters S and Z indicate the typical positions of the particle lines for the 3-stripe and zig-zag phases, respectively.}
\end{figure}

\begin{figure}
\centerline{\psfig{file=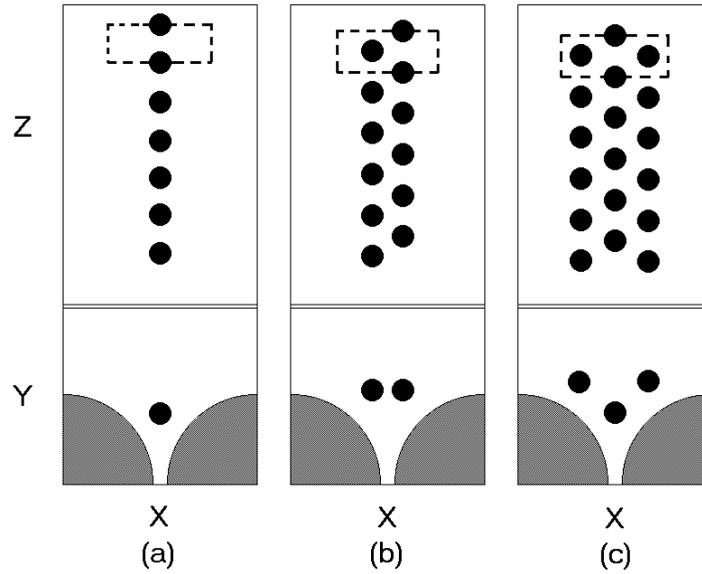, width=11cm}}
\caption{Schematic depiction of three hypothetical phases investigated. Top portion is view looking down on the x-z plane. Bottom view is of the x-y plane. The region marked by the dashed line represents the unit cell. (a)Linear phase, (b) Zig-zag phase, (c) 3-stripe phase.}
\end{figure}

\newpage

\begin{figure}
\centerline{\psfig{file=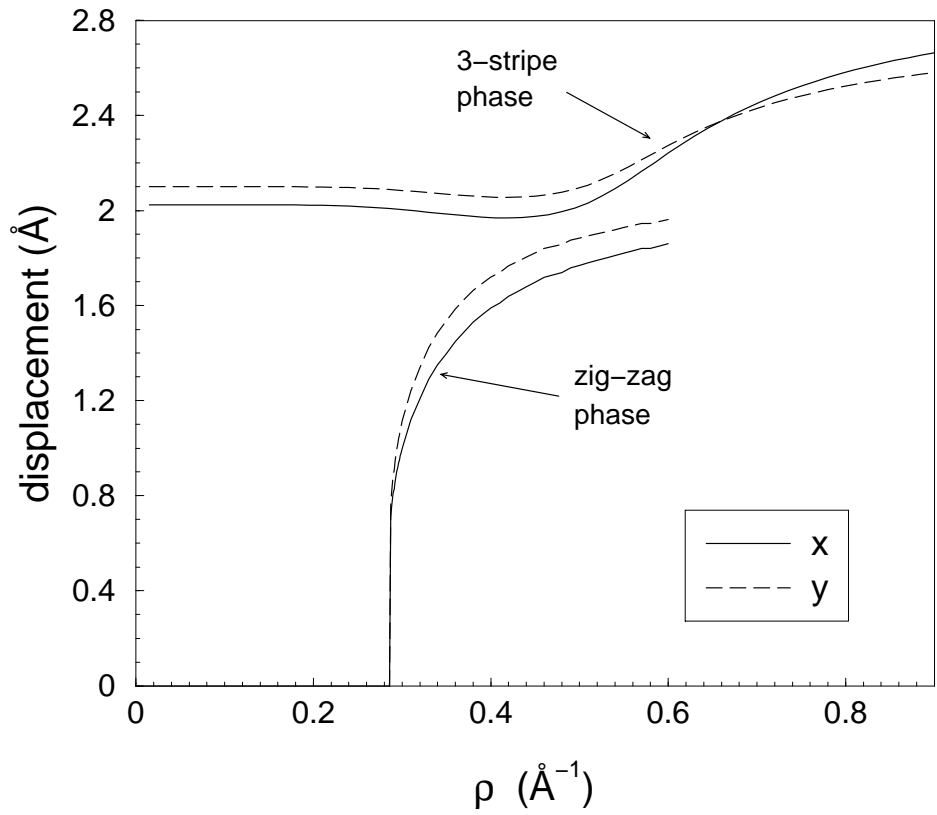, width=11cm,angle=-90}}
\caption{Positions of CH$_4$ molecules in the zig-zag and 3-stripe phases as a function of the linear density. In the zig-zag phase, the two parallel lines are located along the z direction at (-x,y) and (x,y) while in the 3-stripe phase the outer lines are at (-x,y) and (x,y) whereas the center line is at (0,0).}
\end{figure}

\begin{figure}
\centerline{\psfig{file=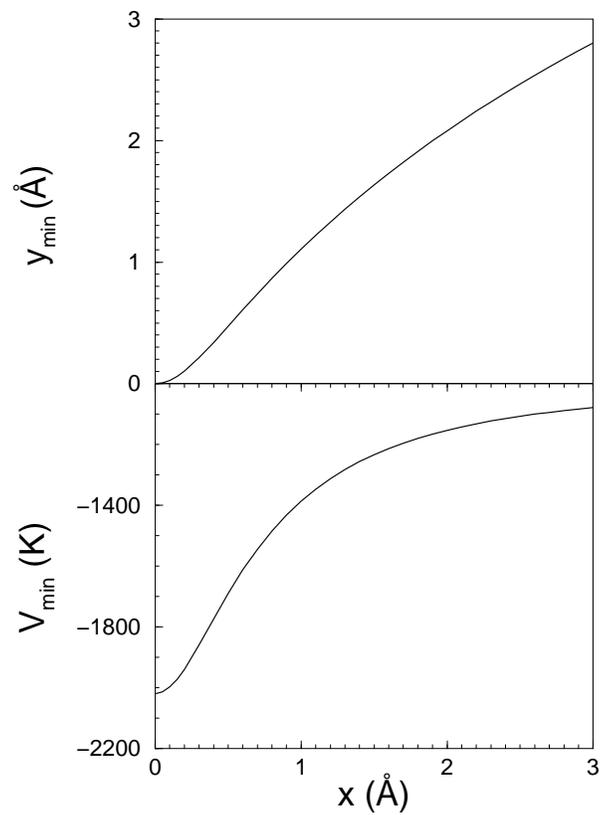, width=11cm,angle=-90}}
\vspace{1.5cm}
\caption{Position of the CH$_4$ potential energy minima (top panel) and energy values at these minima (bottom panel) as a function of assumed x displacement. Note that the asymptotic value of $V_{min}$ as x increases (along the surface, going away from the groove) is the minimum potential energy for graphene, -1070 K.}
\end{figure}

\begin{figure}
\centerline{\psfig{file=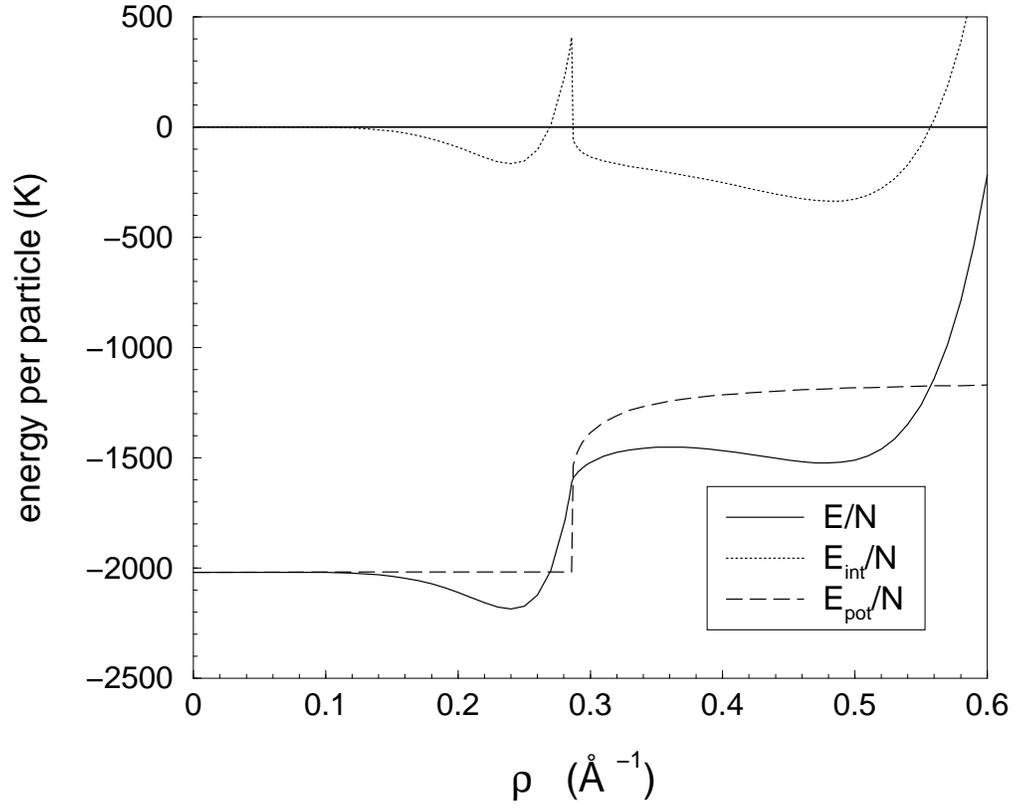, width=11cm,angle=-90}}
\caption{Total energy per CH$_4$ molecule (full line) and the corresponding mutual interaction (dash-dotted line) and adsorption potential (dashed line) energy contributions for the linear and zig-zag phases. Note the jump in the individual contributions when the particles move away from the minimum potential energy position to form a zig-zag phase (near $\rho=0.285$ \AA$^{-1}$).}
\end{figure}

\newpage

\begin{figure}
\centerline{\psfig{file=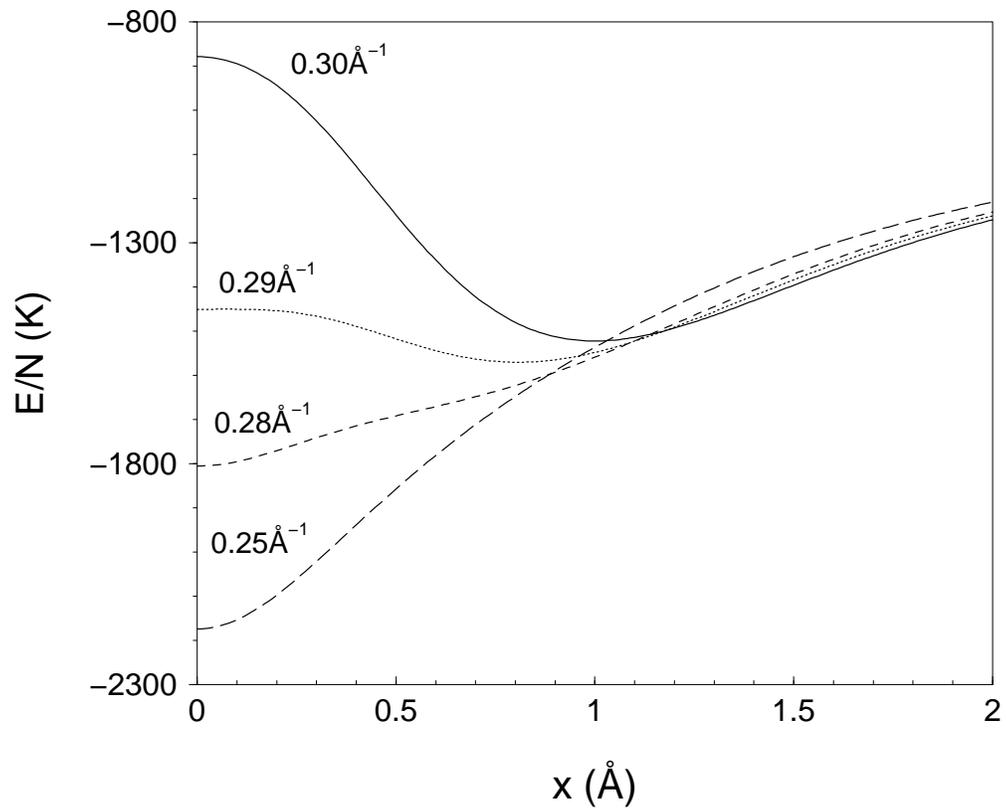, width=11cm,angle=-90}}
\caption{Minimization of the total energy per CH$_4$ particle as a function of the hypotetical displacement x, in search of the zig-zag phase. Each curve corresponds to the labeled value of the linear density. As the density increases (from bottom to top), the first two curves have a minimum at x = 0 (linear phase) while the next two have their minima at a finite value of x (zig-zag phase).}
\end{figure}

\newpage

\begin{figure}
\centerline{\psfig{file=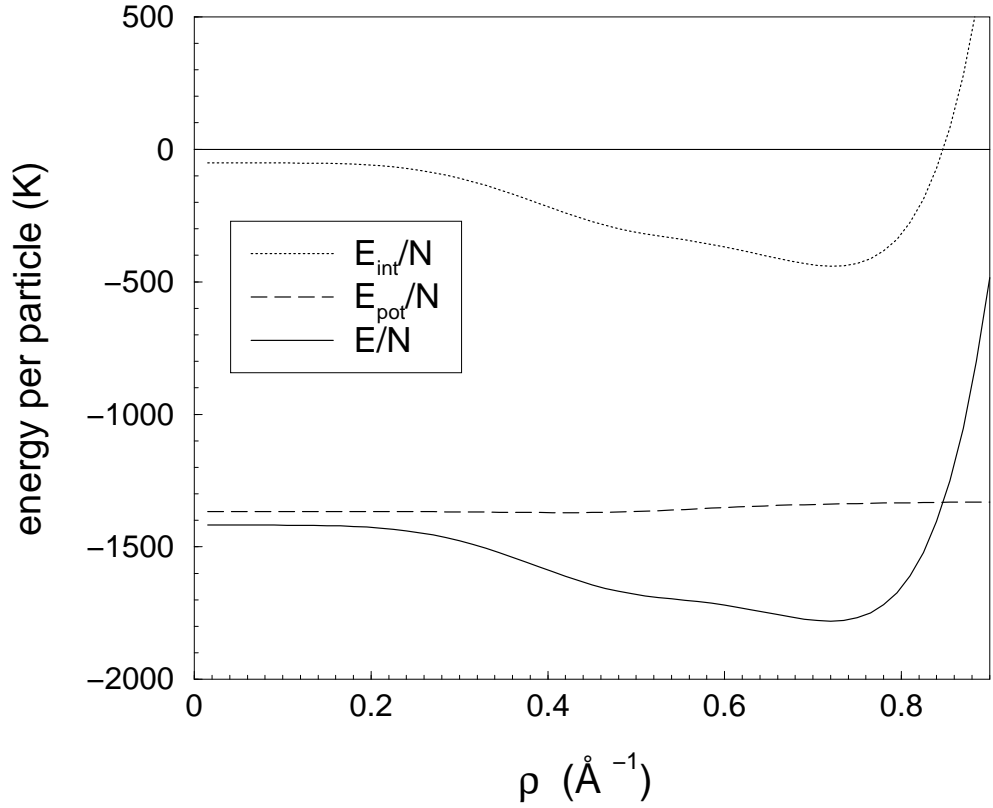, width=11cm,angle=-90}}
\caption{Total energy per CH$_4$ molecule in the 3-stripe phase. Note that the gas-surface potential energy contribution ($E_{pot}$) remains almost constant when the particles choose their favored position for each value of the linear density, while the mutual interaction energy exhibits a minimum near $\rho=0.75$ \AA$^{-1}$.}
\end{figure}

\newpage

\begin{figure}
\centerline{\psfig{file=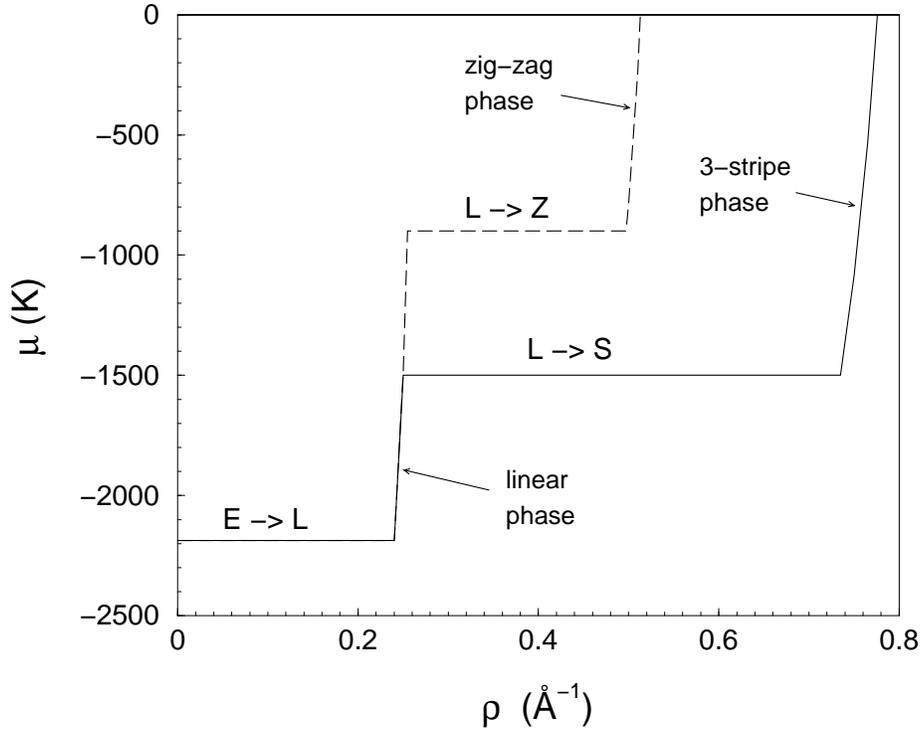, width=10cm,angle=-90}}
\caption{Chemical potential of CH$_4$ at T=0 as a function of 1D density. Below 0.24 \AA $^{-1}$, there occurs coexistence between an empty (E) and a linear (L) phase. For 0.24 \AA$^{-1}< \rho<$ 0.25 \AA$^{-1}$, the linear phase is stable. For 0.25 \AA$^{-1}<\rho<$ 0.735 \AA$^{-1}$, coexistence occurs between the linear and 3-stripe phases. The dashed curve describes the phase behavior of the zig-zag phase, which is not stable.}
\end{figure}

\newpage

\begin{figure}
\centerline{\psfig{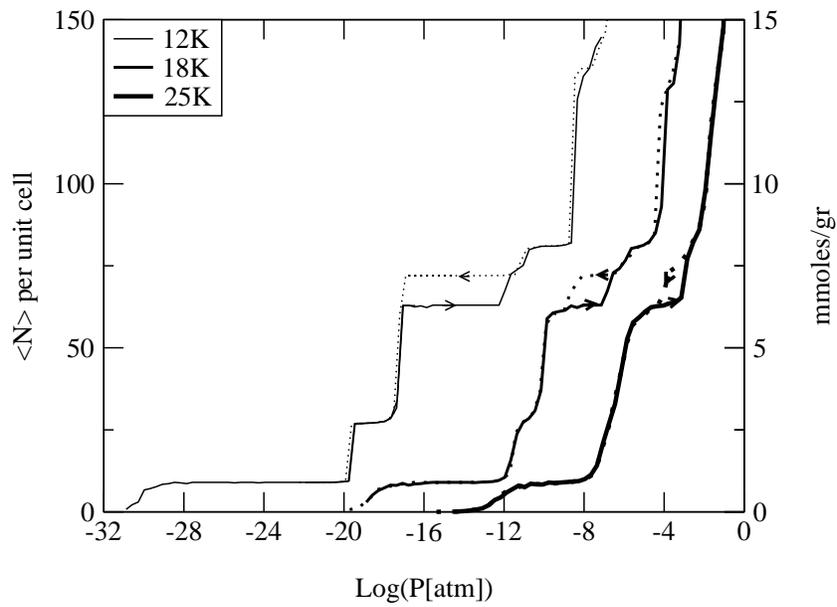}}
\caption{Adsorption and desorption isotherms of Ne at T=12, 18, and 25 K. Note hysteresis near monolayer completion. The unit cell (of length 10 $\sigma_{gg}$) contains a closed-packed line of about 9 atoms. The ordinates' scales are discussed in the text.}
\end{figure}

\newpage

\begin{figure}
\centerline{\psfig{file=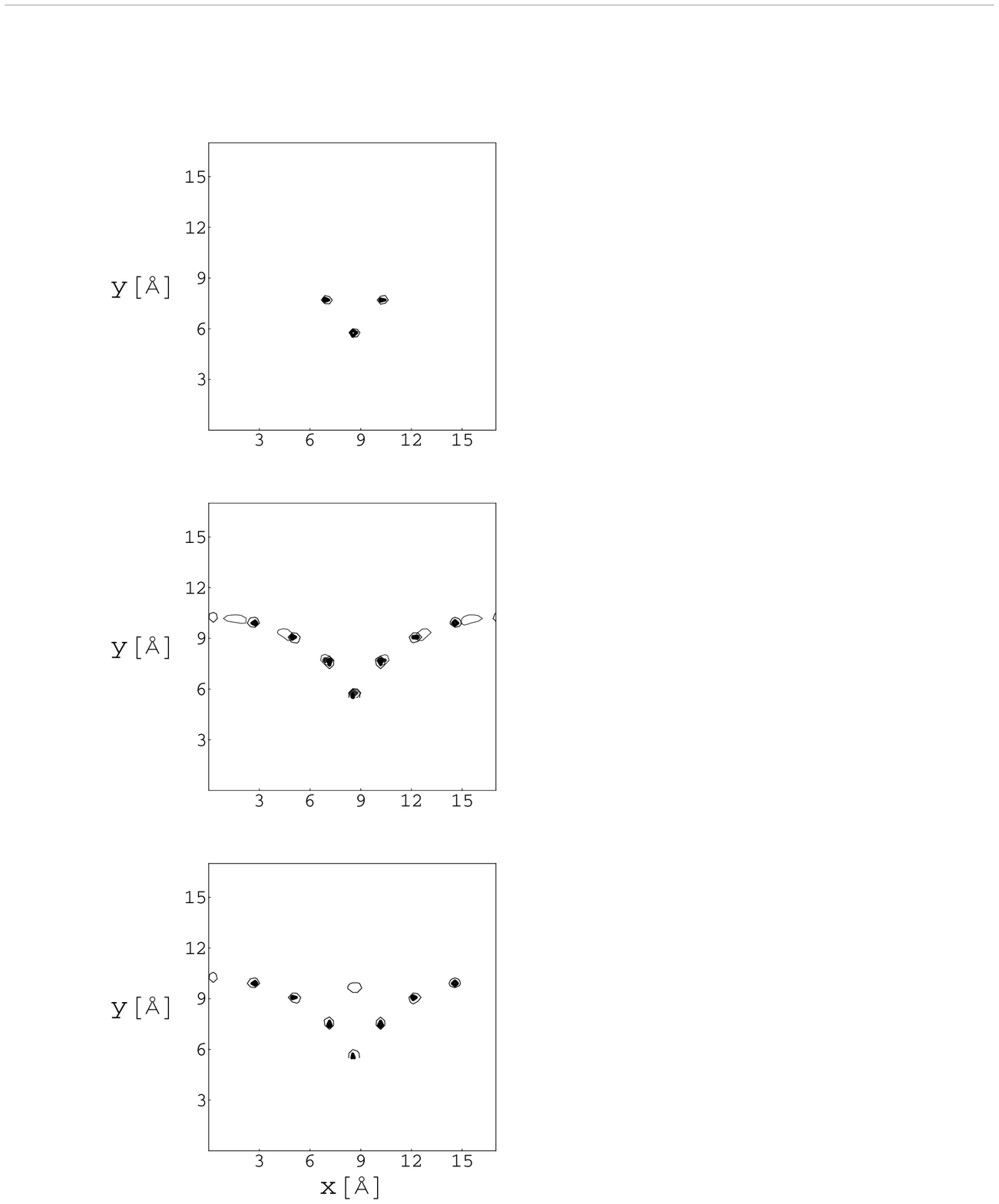, width=15cm}}
\caption{Density contours of Ne at T=12 K projected onto the x-y plane. From top to bottom, the vapor pressures are 10$^{-19}$ atm, 10$^{-16}$ atm and 10$^{-9}$ atm. The middle figure displays superposed density contours of the two competing monolayer phases: 7-stripe (light) and 8-stripe (bold).}
\end{figure}

\newpage

\begin{figure}
\centerline{\psfig{file=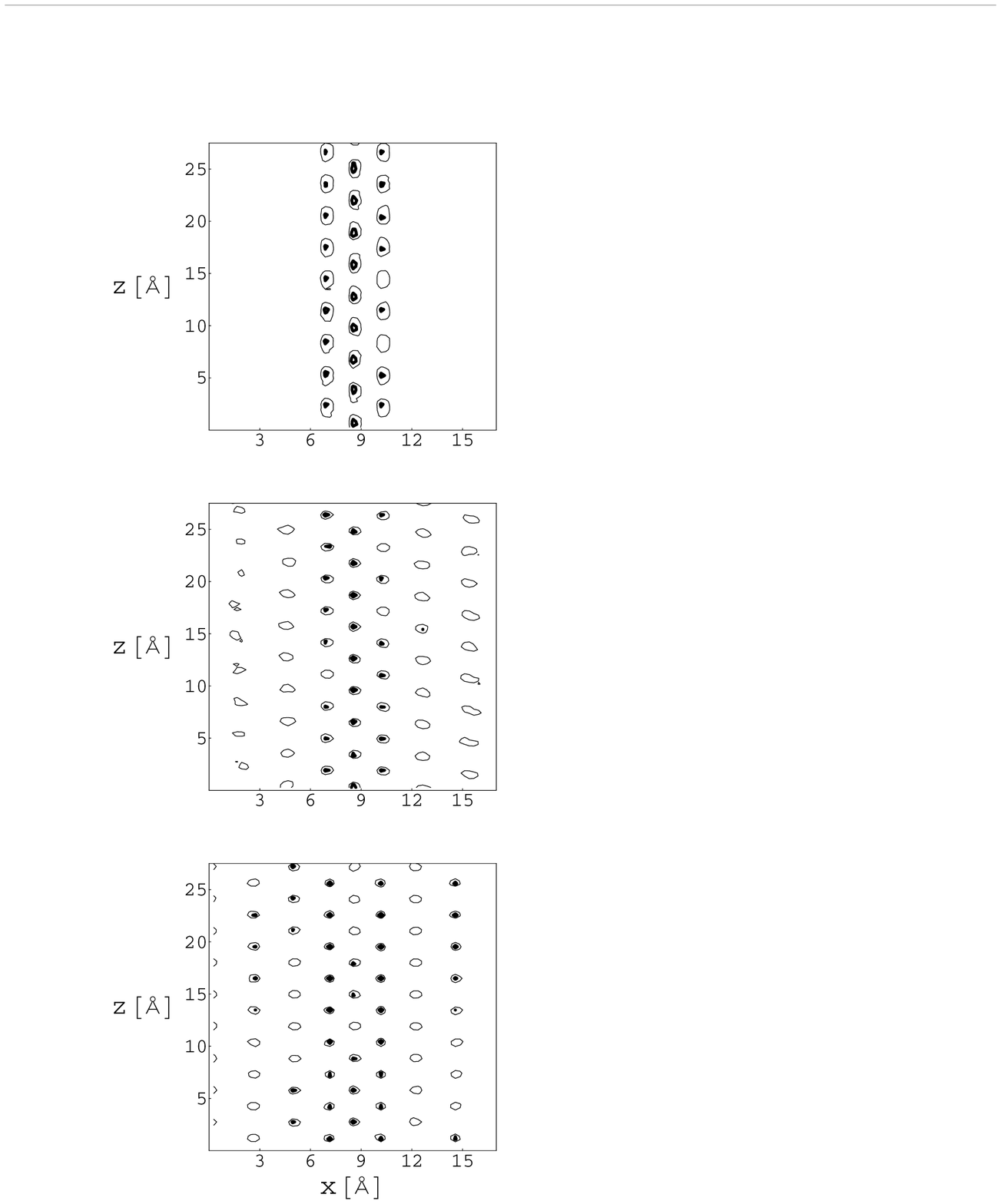, width=15cm}}
\caption{Density contours of Ne at T=12 K projected onto the x-z plane. From top to bottom, 3-stripe phase, P = 10$^{-19}$ atm; 7-stripe phase, P=10$^{-16}$ atm; 8-stripe phase, P=10$^{-16}$ atm.}
\end{figure}

\newpage

\begin{figure}
\centerline{\psfig{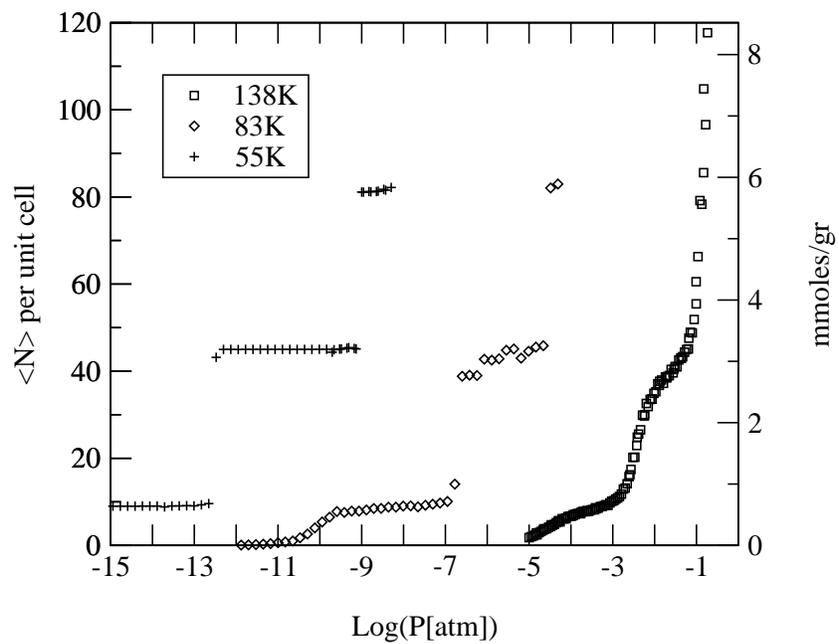}}
\caption{Adsorption isotherms for Xe on the outside of a nanotube bundle at T=
55 K, 83 K, and 138 K.}
\end{figure}

\end{document}